\documentclass[reprint,aps,prl,superscriptaddress]{revtex4-2}
\pdfoutput=1
\usepackage{amsmath,bm}
\usepackage{epsfig}
\usepackage{setspace}
\usepackage{xcolor}
\usepackage{hyperref}

\def\uu{{\bm u}}

\def\REV#1{{\textcolor{black}{#1}}}    % revised text

\begin{document}
\newcommand{\aap}{Astron. Astrophys.}
\newcommand{\araa}{Annu. Rev. Aston.  Astrophys.}
\newcommand{\aapr}{Astron. Astrophys. Rev.}
\newcommand{\mnras}{Mon. Not. R. Astron. Soc.}
\newcommand{\apjl}{Astrophys. J. Lett.}
\newcommand{\apjs}{Astrophys. J. Suppl.}
\newcommand{\jfm}{J. Fluid Mech.}
\newcommand{\jcph}{J. Comput. Phys.}

\newcommand*{\kh}{\color{purple}}
\sloppy
\def\uu{{\bm u}}
\def\bb{{\bm b}}
\def\rr{{\bm r}}
\def\kk{{\bm k}}

\title{Statistical signatures of interstellar turbulence in dust polarization maps}

\author{Ka Wai Ho}\email{kawaiho@kitp.ucsb.edu}
\affiliation{Department of Astronomy, University of Wisconsin-Madison, WI 53715, USA}
\affiliation{Theoretical Division, Los Alamos National Laboratory, Los Alamos, NM 87545, USA}
\affiliation{Kavli Institute for Theoretical Physics, University of California, Santa Barbara, CA 93106, USA}
\author{Ka Ho Yuen}\email{Present address: 
Nanjing University, Suzhou, Jiangsu Province, PRC; kyuen@nju.edu.cn}
\affiliation{Theoretical Division, Los Alamos National Laboratory, Los Alamos, NM 87545, USA}
\author{Raphael Flauger} \email{flauger@ucsd.edu}
\author{Alexei G. Kritsuk}\email{akritsuk@ucsd.edu}
\affiliation{Physics Department, University of California, San Diego, La Jolla, CA 92093, USA }

\begin{abstract} 

We present results from a high-resolution interstellar turbulence simulation and show that it closely reproduces recent {\em Planck} measurements. Our model captures the scaling of $EE$ and $BB$ spectra, and the $EE/BB$ ratio in the inertial range. The probability density function of the dust polarization fraction is also consistent with observations. The $TE$ cross-correlation is in broad agreement with the {\em Planck} sky. This simulation provides new insights into the physical origins of the observed $E/B$ asymmetry and positive $TE$ signal, facilitating the development of advanced Galactic dust emission models for current and future cosmic microwave background experiments.
\end{abstract}


\maketitle

\paragraph{Introduction.---}

Observations of the cosmic microwave background (CMB) have allowed us to determine key properties of our Universe, including its age, geometry, and composition. These observations have also tightly constrained the statistical properties of the primordial density fluctuations that seeded the large-scale structure of the Universe. Precision measurements of the CMB polarization anisotropies have the potential to reveal deep insights into the physical processes that generated these primordial perturbations \cite{kamionkowski.16}. 

However, the CMB is far from the only source of polarized radiation in the microwave sky. The Galactic magnetic field threads the interstellar medium (ISM) and aligns dust grains, leading to polarized emission from thermal dust. In addition, relativistic electrons spiraling around magnetic field lines emit highly polarized synchrotron radiation. These Galactic sources of polarized emission are orders of magnitude brighter than the inflationary signal at all frequencies, even in the cleanest patches of the sky \cite{dickinson16}. As a consequence, detecting the primordial signal requires exquisite control over Galactic foregrounds. To achieve this, experiments rely on multifrequency observations. Optimizing the frequency distribution during the design stage, accurately estimating covariance matrices, and judging the robustness of a possible detection all require an understanding of or the ability to simulate the ISM.

The ISM is a complex magnetized mix of neutral and ionized gas, dust, cosmic rays, and radiation, all coupled through mass, energy, and momentum exchange. The conducting fluid component, including neutral and ionized hydrogen, is thermally unstable in certain density and temperature regimes \cite{field65} and tends to split into two or more stable thermal phases \cite{goldsmith..69,zeldovich.69,wolfire...03}. The ISM is also constantly energized by star formation feedback, including supernova explosions, which supplements the incoming energy cascade from global scales of the Galactic disk, keeping the fluid turbulent \cite{maclow.04,hennebelle.12,klessen.16,colman.........22,elmegreen24}. Understanding the structure of this highly compressible multiphase magnetized turbulence is a challenge because of the multiscale nature of involved nonlinear interactions. Hence, the focus of recent studies was mostly on numerical experiments. 

In this Letter we analyze synthetic dust polarization maps derived from a new simulation of magnetized multiphase ISM turbulence complementary to the set of cases described in Refs.~\cite{kritsuk..17,kritsuk..18}.

\paragraph{Numerical model.---}
 This new higher-resolution model solves the equations of ideal magnetohydrodynamics (MHD), augmented by terms describing external forcing and a generalized volumetric energy source; see Eqs.  (1a)--(1d) in Ref.~\cite{kritsuk..17}. We use the same prescription for cooling and heating processes, and the same domain size of $L=200$~pc. The mean H{\sc i} number density, $n_0=1$~cm$^{-3}$, corresponds to a mean column density $N_{\rm H}\approx6\times10^{20}$~cm$^{-2}$. This is close to the mean column density in the largest of the six nested regions studied by the {\em Planck} collaboration in Ref.~\cite{pr3.20}, which covers 71\% of the sky. The mean magnetic field strength is $b_0=5\;\mu$G and the target rms velocity is $u_{\rm 0,rms}=20$~km/s, resulting in the nominal Alfv\'en Mach number ${\cal M}_{A,0}\equiv u_{\rm 0,rms}/(b_0/\sqrt{4\pi\rho_0})\approx2.1$ and dynamical (large-eddy turnover) time $\tau_{\rm d}\equiv L/(2u_{\rm 0,rms})\approx5$~Myr. To induce and support turbulence, we used purely solenoidal \cite{kritsuk...10} large-scale Ornstein-Uhlenbeck forcing with correlation time $t_{\rm corr}=2$~Myr (cf. $t_{\rm corr}=\infty$ in Ref.~\cite{kritsuk..18}). The model was evolved for a total of 60~Myr, and turbulence can be deemed fully developed after the first $3\tau_{\rm d}=15$~Myr. The resulting dataset includes 35 full flow snapshots collected at $t\geq3\tau_{\rm d}$, which were used for statistical averages in the stationary state.
Finally, the nominal grid resolution in this model is $2112^3$, corresponding to the linear grid spacing of $\Delta x\sim0.095$~pc \footnote{Note the new simulation employs methods with piecewise-linear (PLM, \cite{mignone14}) reconstruction, while previous lower resolution models used piecewise-parabolic reconstruction (PPML, \cite{ustyugov...09}). Both codes used the HLLD Riemann solver.}. The simulation does not include explicit viscosity and magnetic diffusivity. We expect numerical dissipation to {\em directly} affect scales below $\sim32\Delta x$ \cite{sytine....00}. 
\begin{figure*}
	\centering
	\includegraphics[scale=.46]{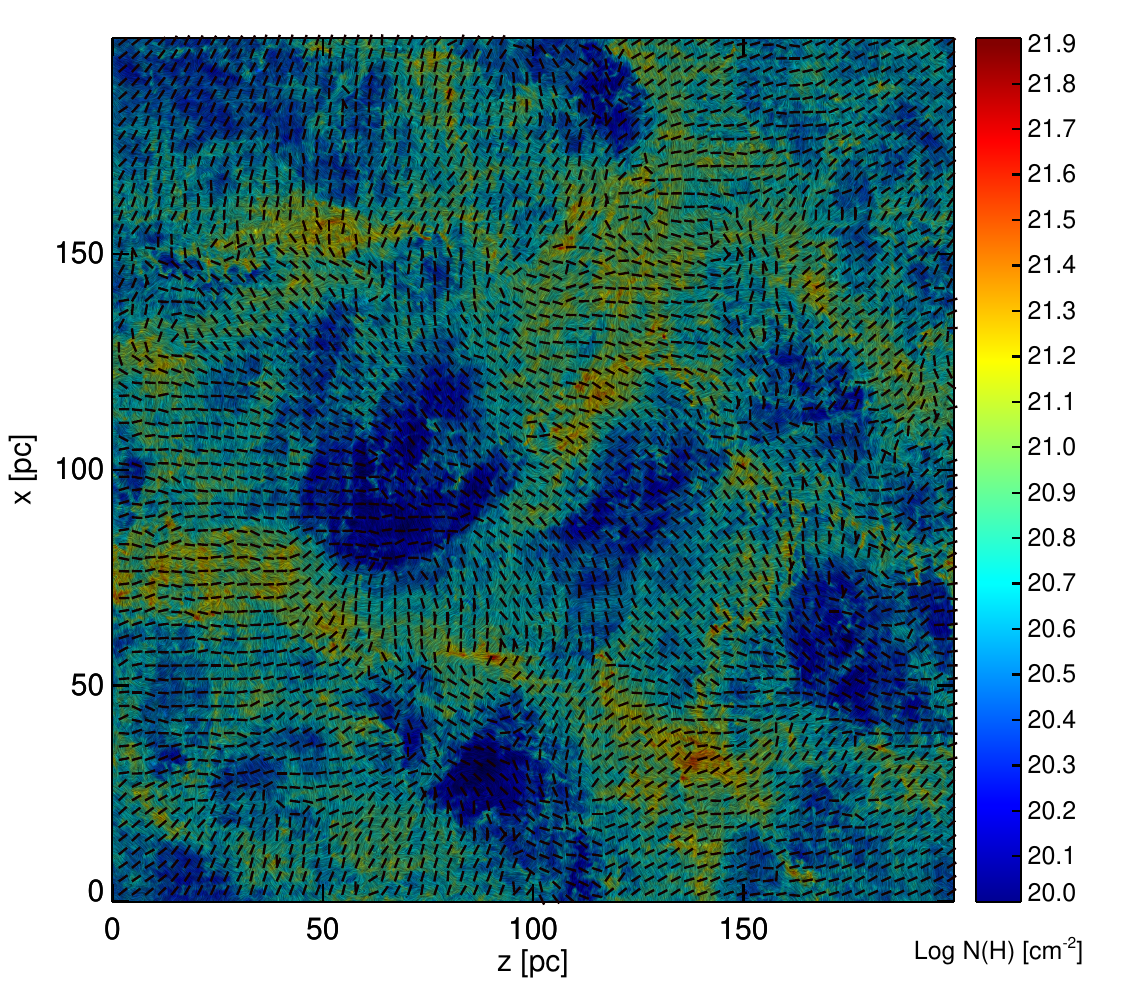}
	\includegraphics[scale=.46]{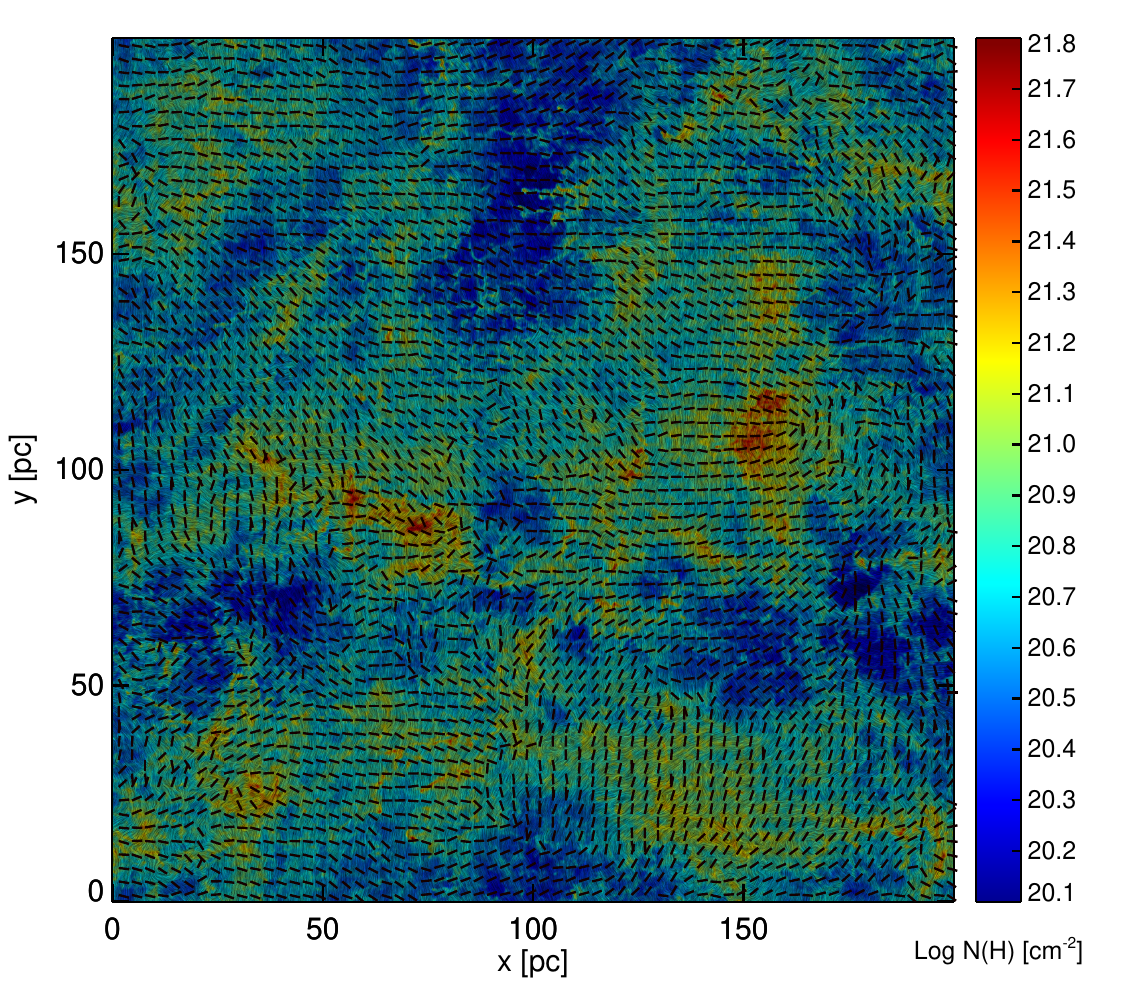}
	\caption{Sample synthetic polarization maps, showing projections parallel (left) and perpendicular (right) to the mean field $\bm b_0$. The drapery texture generated using the LIC technique \cite{cabral.93} shows the POS magnetic field structure. Pseudovectors indicate the polarization direction (predominantly perpendicular to the field). Color shows the intensity in units of H{\sc i} column density. These maps are built on full-resolution $2112^3$ numerical data smoothed with a low-pass box-car filter of length 5 voxels. 
	}
	\label{map}
\end{figure*}
Anticipating {\em indirect} effects of numerical dissipation \footnote{E.g., the bottleneck phenomenon \cite{falkovich94}} and convolution effects associated with the projection operation,  we conservatively presume the inertial range would be within $\ell{\rm (pc)}\in[6.3,63]$ or, equivalently, in a range of wave numbers $\log_{10}(k/k_{\rm min})\in[0.5,1.5]$. This is a significant improvement over previous MHD models \cite{kritsuk..18}, which barely resolved the inertial range. We estimate that with $4\times$ higher linear grid resolution the effective Reynolds number, $\Re_{\rm eff}$, is $\sim6\times$ higher, making the model more realistic and offering an opportunity to compare model predictions with {\em Planck} measurements for a wider set of statistics, now additionally including $EE$, $TB$, $TE$, and $r_{TE}$ correlations, $\alpha^{EE}$ and $\alpha^{TE}$ spectral slopes, $TE/EE$ spectral ratio, as well as  polarization fraction PDF.

\paragraph{Statistics of developed turbulence.---}
Our simulation of multiphase MHD turbulence covers a range of scales $0.1-200$~pc and naturally includes a variety of coexisting MHD regimes of ISM turbulence. Under the local Galactic conditions, nonlinear relaxation leads to sub-Alfv\'enic turbulence in the space-filling  warm and thermally unstable phases (99\% by volume), while the cold neutral medium  and molecular gas (comprising $\sim$25\% by mass) are super-Alfv\'enic; see Table~\ref{equi}.

In terms of the nominal Alfv\'en Mach number, this model is somewhat similar to the low-resolution low-density case \textsf{D} from Ref.~\cite{kritsuk..17} ($n_0=2$~cm$^{-3}$, ${\cal M}_{A,0}=2.6$) and falls in between the high-density ($n_0=5$~cm$^{-3}$) cases \textsf{A} (${\cal M}_{A,0}=1.3$) and \textsf{B} (${\cal M}_{A,0}=4.0$) from Ref.~ \cite{kritsuk..18}. However, Alfv\'en Mach numbers defined through rms Alfv\'en velocity, ${\cal M}_{\rm a}\equiv u_{\rm rms}/v_{\rm a,rms}$, for these three cases are quite different. Indeed, in our high-resolution case ${\cal M}_{\rm a}=0.4$, while in cases \textsf{A} and \textsf{B} ${\cal M}_{\rm a}=1.0$ and 1.4, respectively. Meanwhile, volume-weighted rms sonic Mach numbers are ${\cal M}_{\rm s}=3.4$, 4.9, and 5.4, respectively.  Thus, in all cases turbulence is supersonic. A combination of supersonic velocity fluctuations at the energy containing scale, enhanced compressibility due to effects of thermal instability, and four-times refined $2112^3$ grid yields standard deviation of the density $\sigma_{\rho}/\rho_0=4.6$, which is significantly higher than $\sigma_{\rho}/\rho_0=3.7$ and 3.5 measured in cases \textsf{A} and \textsf{B}, respectively.

Due to a substantially lower mean H{\sc i} density, our high-resolution case is dominated by the warm phase, which occupies 64\% of the volume. This is in contrast to cases \textsf{A} and \textsf{B} from  Ref.~\cite{kritsuk..18}, where the warm phase fills only 25\% and the unstable phase fills up to 70\%. Likewise, the cold phase fills less that 1\% of the volume and about 7\% in cases \textsf{A} and \textsf{B}. These differences in phase volume filling factors ${\cal F}_{\rm v}$ explain both overall lower sonic and Alfv\'en Mach numbers here compared to cases \textsf{A} and \textsf{B}.

\begin{table}[b]
\caption{\label{par}Characteristics of statistical equilibrium: volume (${\cal F}_{\rm v}$) and mass (${\cal F}_{\rm m}$) fractions of warm (W: $T>5250$~K), unstable (U: $184<T<5250$~K), and cold (C: $T<184$~K) thermal gas phases, volume-averaged sonic (${\cal M}_{\rm s}$) and Alfv\'en  (${\cal M}_{\rm a}$) Mach numbers,  thermal and turbulent plasma-$\beta$, relative rms field fluctuations $b'_{\rm rms}/b_0$, and corresponding phase-average quantities.}
\footnotesize
\begin{tabular}{@{}cccccccc}
\hline
\hline
Phase & ${\cal F}_{\rm v}$ & ${\cal F}_{\rm m}$ & ${\cal M}_{\rm s}$ & ${\cal M}_{\rm a}$ & $\beta_{\rm th}$&  $\beta_{\rm turb}$ & $b'_{\rm rms}/b_0$ \\\hline
C & 0.01 & 0.25 & 17 & 3.1 & 0.08 & 41.2 & 2.6 \\
U & 0.35 & 0.52 & 3.8 & 0.5 & 0.13 & 3.00 & 2.2 \\
W & 0.64 & 0.23 & 2.4 & 0.4 & 0.09 & 0.84 & 2.1 \\
All & 1.00 & 1.00 & 3.4 & 0.4 & 0.10 & 1.92 & 2.1 \\
 \hline
\end{tabular}
\label{equi}
\end{table}

\begin{figure*}
	\centering
	\includegraphics[scale=.65]{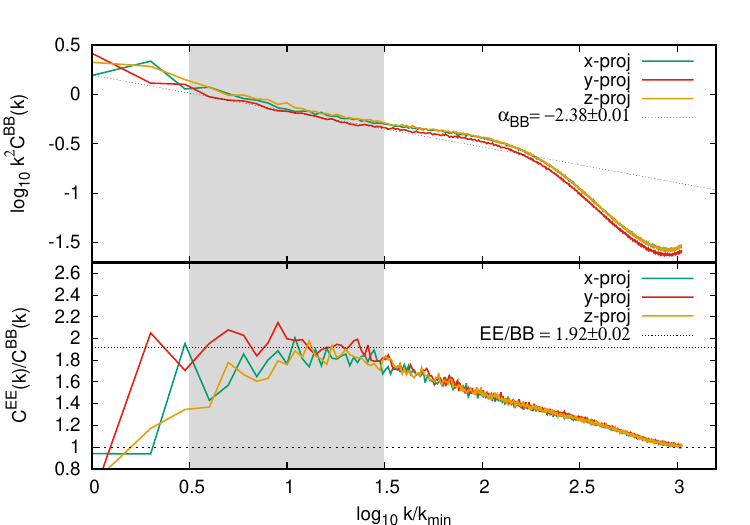}
	\includegraphics[scale=.65]{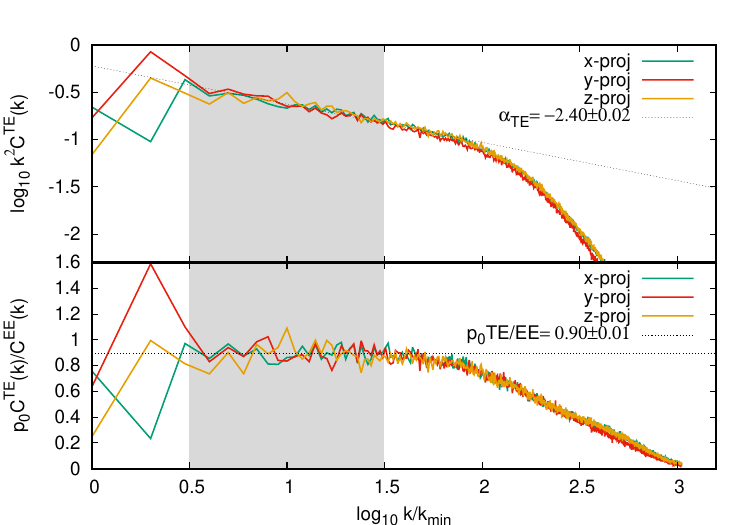}	
	\caption{ Compensated time-average $BB$ (left panel, top) and $TE$ (right panel, top) (co)spectra in arbitrary units, $EE/BB$ and $TE/EE$ spectral ratios (bottom). Dotted lines in the top plots show least-squares fits for projection along the mean field $\bm b_0$ in a range of $\log_{10}(k/k_{\rm min})\in[0.5,1.5]$. Horizontal dotted lines in the bottom plots indicate similar fits for the spectral ratios.
		}
	\label{ebpow}
\end{figure*}

In terms of plasma beta, $\beta_{\rm th}\equiv\langle p/(b^2/8\pi)\rangle=\langle2c_{\rm s}^2/\gamma v_A^2\rangle=0.1$, our new model is similar to case \textsf{A} and shows lower than average values in the cold and warm phases, while in the thermally unstable gas $\beta_{\rm th}$ is slightly higher than average. In terms of turbulent plasma beta, $\beta_{\rm turb}\equiv\langle \rho u^2/(b^2/8\pi)\rangle=2\langle K/M\rangle=1.9$, the new model is also close to case \textsf{A} of Ref.~\cite{kritsuk..17}. (Here, $K$ and $M$ are kinetic and magnetic energy densities.) 
Note that the cold phase shows extreme $\beta_{\rm turb}\approx40$ or $\langle K/M\rangle\approx20$, which implies strongly kinetically dominant regime. The thermally unstable gas is weakly kinetically dominated, while the volume-filling warm phase is mildly magnetically dominated (see Table~\ref{equi}). Finally, the relative rms fluctuations of magnetic field strength are $\sim 2$, which places our new model right in between cases \textsf{A} and \textsf{B}. As expected, the fluctuations are larger in the cold gas due to enhanced compressibility. It is worth noting that the prediction of super-Alfv\'enic nature of turbulence in the cold/molecular phase made in Ref.~\cite{kritsuk..11} is now validated with our high-resolution simulation.

Overall, our $2112^3$ model together with cases \textsf{A} and \textsf{B} from Ref.~\cite{kritsuk..18} covers a broad range of multiphase turbulence regimes with ${\cal M}_{A,0}\in[1.3,4.0]$, ${\cal M}_{\rm a}\in[0.4,1.4]$, ${\cal M}_{\rm s}\in[3.4,5.4]$, $\beta_{\rm th}\in[0.1,0.4]$, and $\beta_{\rm turb}\in[2,4]$. It is remarkable that all these cases with isotropic solenoidal forcing reproduce the $E$- and $B$-mode spectra and spectral ratios observed by {\em Planck} reasonably well, cf.~Refs.~\cite{cho.23,stalpes..24}.

\paragraph{Polarization maps.---}
The maps are constructed for three line-of-sight directions, coinciding with the principal coordinate axes and define projected quantities as functions of position $(x,y)$, $(y,z)$, and $(z,x)$, assuming the medium at the wavelengths of interest is optically thin. For a projection along the $z$ direction the intensity $I(\rr)$ and the Stokes parameters $Q(\rr)$ and $U(\rr)$ are computed as follows:
$ I=\int\rho dz$,
$Q=p_0\int\rho{(b_y^2-b_x^2)}/{b^2}dz$, and 
$U=-2p_0\int\rho {b_xb_y}/{b^2}dz$,
where $p_0$ is the polarization fraction, $\bm b\equiv\bm b_0+\bm b^{\prime}=(b_x, b_y, b_z)$,  and dust grains are assumed to be perfectly aligned with the magnetic field. As in Ref.~\cite{kritsuk..18}, we introduce masking and replace the density weighing factor $\rho$ under the integral with $\rho\theta(\rho_t-\rho)$
controlled by the threshold density $\rho_t$, where $\theta(\rho)$ is the Heaviside step function. We will discuss the need for and effects of masking in detail below. The polarized intensity is then given by $P=\sqrt{Q^2+U^2}$, while the polarization angle is $\psi_b=\arctan\left({U}/{Q}\right)/2$. With these definitions, one can compute synthetic maps of $I(\rr)$, $Q(\rr)$, $U(\rr)$, and $P(\rr)$. One can also compute the actual plane-of-sky (POS) magnetic field for this same projection $\tilde{\bm b}(\bm r)=(\tilde{b}_x,\tilde{b}_y)$.

As an illustration, Fig.~\ref{map} shows two sample maps for the $2112^3$ model, using projections along axes parallel (left) and perpendicular (right) to the mean magnetic field $\bm b_0=(0,b_0,0)$. As expected, the thermal dust polarization direction is mostly perpendicular to the direction of POS magnetic field $\tilde{\bm b}$ shown by the drapery pattern.  
We note, however, that due to a smaller degree of anisotropy in this model the two maps look qualitatively similar, unlike in case \textsf{A} where a two times  stronger $b_0$ results in a significantly more regular and anisotropic structure of $\tilde{\bm b}$ in the maps with the mean magnetic field in the POS.
%However, the two maps look qualitatively different overall, with a more regular and anisotropic structure of $\tilde{\bm b}$ in the right panel, where the mean magnetic field lies in the POS.

\paragraph{The $EE$, $BB$ spectra and $E/B$ asymmetry.---}
For each polarization map we construct maps of $E$ and $B$ modes, which are computed using the Fourier transforms of $Q(\rr)$ and $U(\rr)$ according to Eqs. (4) and (5) from Ref.~\cite{kritsuk..18}.
In turn, the $E$- and $B$-mode spectra are defined by $C^{BB}(k)=\langle|\widehat{B}(\kk)|^2\rangle$ and $C^{EE}(k)=\langle|\widehat{E}(\kk)|^2\rangle$, taking an average $\langle\cdot\rangle$ over all wave vectors $\kk=(k_1,k_2)$ satisfying $|\kk|=k$. Figure~\ref{ebpow} (left panel) shows the $BB$ spectra (top), as well as $EE/BB$ ratios (bottom) for all projections, using averaging over 35 flow realizations \footnote{To accurately measure statistics of polarized foregrounds in our model for stationary and homogeneous ISM turbulence, we use ensemble averages for a reasonably large set of high-resolution synthetic maps sampled over a period of $9\tau_{\rm d}=45$~Myr. We then rely on the ergodic hypothesis (see, e.g.~\cite{monin.65}) to justify the comparison with observational data for LR71 ---the largest region in the {\em Planck} sky, for which accuracy is achieved through extensive space averaging}.

\begin{figure*}
	\centering
	\includegraphics[scale=.65]{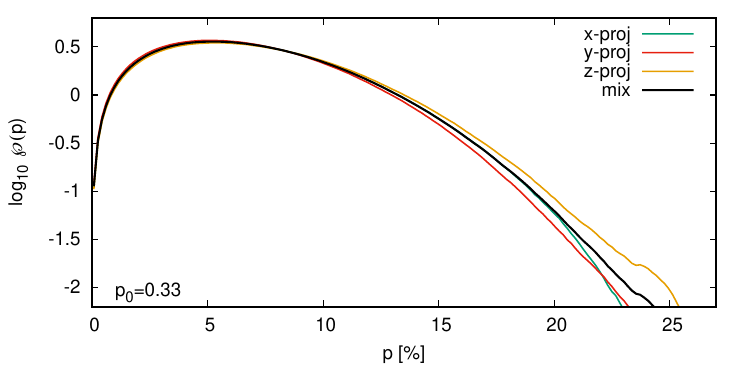}
	\includegraphics[scale=.65]{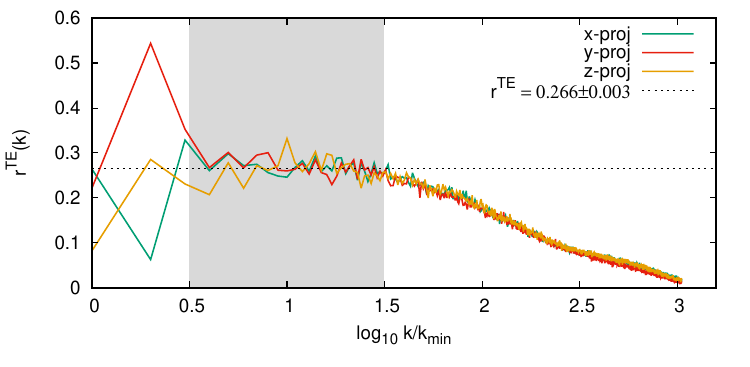}	
	\caption{Polarization fraction PDF for projections along the principal axes $x$, $y$, and $z$ as well as the mean PDF for an equal mix of all three projections (left panel).
	Correlation coefficient $r_{TE}(k)$ (right panel). Horizontal dotted line indicates least-squares fit to $r_{TE}(k)$ for projection along $\bm b_0$ in a range of $\log_{10}(k/k_{\rm min})\in[0.5,1.5]$.
	}
	\label{rte}
\end{figure*}

The calculations assume a density mask with threshold value $\rho_t=40\rho_0$. 
The masked voxels comprise  a small fraction, $\sim0.17$\%, of the volume. With moderate magnetization of our model, the spectra bear a very mild signature of large-scale turbulence anisotropy, with the $y$-projection spectrum carrying less power than its $x$ and $z$ counterparts.

Naturally, the spectra are subject to the usual resolution constraints implied by the numerics; hence, the spectral interval of interest here is limited to the range of wave numbers $\log k/k_{\rm min}\in[0.5,1.5]$. The slope of the $BB$ spectrum in this range, $\alpha_{BB}=-2.38\pm0.01$, is close to $-2.54\pm0.02$ measured by {\em Planck} for the LR71 sky region. The slope of the $EE$ spectrum in the simulation, $\alpha_{EE}=-2.41\pm0.01$, is also close to {\em Planck's} measurement of $\alpha_{EE}=-2.42\pm0.02$ \footnote{It is worth noting that formal standard deviations provided here and below for the model-based spectral indices (and spectral ratios) only reflect the quality of the least-squares fit of the average spectrum for the projection along the mean field in the specified interval of wave numbers. The accuracy of these spectral indices depends on: (i) the chosen interval of wave numbers, which we deem as inertial range; (ii) the unknown orientation of the mean magnetic field with respect to the line of sight; (iii) the variation of slope between individual flow snapshots; (iv) the number of correlation times in the time-averaging interval; etc. Therefore, it is not hard to imagine that the actual standard deviation of our measurement is of the order of 0.05 or larger.}.

The $E/B$ asymmetry discovered by {\em Planck} is also well captured by the simulation. The least-squares fit to $EE/BB$ spectral ratio for the projection along the mean magnetic field yields $EE/BB=1.92\pm0.02$, while the perpendicular projections show somewhat lower ratios. Note that these estimates are more robust compared to those in Ref.~\cite{kritsuk..18} due to higher resolution of the simulation used here, and they are also in better agreement with {\em Planck} PR3 measurements \cite{pr3.20} in parts of the sky with comparable column densities.

\paragraph{$TE$ correlation.---}
Our simulation exhibits a significant positive temperature-$E$-mode ($TE$) correlation (Fig.~\ref{ebpow}, right panel) with a slope $\alpha_{TE}=-2.40\pm0.02$. This is slightly shallower than {\em Planck's} measurement of $-2.50\pm0.02$ for the LR71 sky region over the multipole range $\ell\in[40,600]$. By comparing the ratio of the $TE$ and $EE$ power spectra $p_0TE/EE=0.90\pm0.01$ measured in the simulation with {\em Planck's} measurement of $TE/EE=2.76\pm0.05$, one can estimate the value of polarization fraction to be $p_0\approx0.33$. Using this value of $p_0$, we calculated the probability density function (PDF) of polarization fraction. This is shown in the left panel of Fig.~\ref{rte} and agrees quite well with that from {\em Planck} \cite{planckXIX.15,planckXLIV.16,planckXII.20} both in terms of its overall shape and effective width. Note that our measurement of $p_0TE/EE$ is quite robust as this spectral ratio is approximately constant across the well-defined inertial range. The spectral ratios involving $B$ modes are generally less well behaved, as can be seen in Fig.~\ref{ebpow}, left panel. In particular, due to comparatively high noise levels in the relevant spectra based on the simulation data, we cannot detect the positive $TB$ dust signal reported in Refs.~\cite{planckXXX.16,pr3.20}.

We also combined the $TE$, $TT$, and $EE$ spectra to compute the dimensionless correlation coefficient $r_{TE}(k)\equiv C^{TE}(k)/\sqrt{C^{TT}(k)C^{EE}(k)}$ presented in Fig.~\ref{rte}, right panel. The coefficient shows no systematic dependence on the wave number or on the projection direction in the shaded interval of wave numbers and the least-squares fit in that range yields $r_{TE}=0.266\pm0.003$. This $r_{TE}$ value is $\sim25$\% lower than the weighted mean of {\em Planck's} measurements for six nested sky regions and multipoles $\ell\in(5,100)$, $r_{TE}=0.357\pm0.003$ \cite{pr3.20}.

\begin{figure}
	\centering
	\includegraphics[scale=.65]{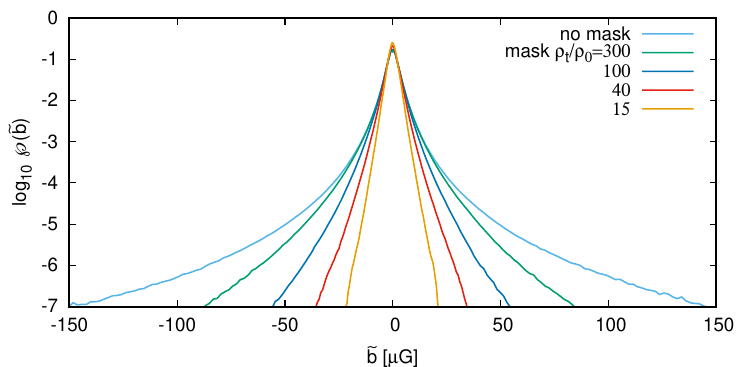}
	\caption{Effects of masking on the PDF of POS magnetic field fluctuations $\tilde{b}$.}
	\label{bpos}
\end{figure}

\paragraph{Masking.---}
\REV{Compressible MHD turbulence simulations with kinetic energy injection, using large-scale solenoidal accelerations, exhibit specific spatio-temporal intermittency manifested in fat stretched-exponential tails in the 
PDFs of magnetic field fluctuations (e.g., Section 3.10 and Fig.~13 in Ref.~\cite{kritsuk..17}). These strongly non-Gaussian distributions were shown to originate primarily in the cold and dense filamentary structures in simulations with relatively low $\Re_{\rm eff}$ (Fig.~14 in Ref.~\cite{kritsuk..17}). Since this spurious intermittency caused by artificial pumping directly affects the statistics derived from synthetic polarization maps, Ref.~\cite{kritsuk..18} introduced density masking (controlled by a threshold $\rho_{\rm t}$) that effectively cuts off the fat tails in the PDF of POS magnetic field strength, $\wp(\tilde{b})$, leaving behind a compact exponential distribution. 
With a higher resolution model here, we support the conjecture made in \cite{kritsuk..18} that at sufficiently high $\Re_{\rm eff}$ the need for masking would be eliminated. We illustrate our choice of the masking threshold $\rho_t/\rho_0=40$ with Fig.~\ref{bpos}, which shows how $\wp(\tilde{b})$ depends on $\rho_{\rm t}$.}

By applying the density mask with $\rho_t/\rho_0=40$ \REV{(compare to $\rho_t/\rho_0=14$ \cite{kritsuk..18})}, we remove a tiny part of the cold (presumably molecular), shocked supersonic and super-Alfv\'enic material packed into dense filaments with the most erratic structure of the magnetic field that would otherwise contribute to the line-of-sight convolutions due to density weighting in the expressions for $I$, $Q$, and $U$. The neglected 0.17\% \REV{(compare to 0.8\% \cite{kritsuk..18})} of voxels would then noticeably randomize synthetic dust-polarization maps, reducing the $E/B$ ratio from 1.9 to $\sim(1.5-1.6)$ and making the $E$- and $B$-mode spectra shallower, with power law indices $-2.15$ and $-2.02$, respectively. 
With the $2112^3$ grid, we reduced the masked volume fraction \REV{by a factor of 4.75} compared to the $512^3$ cases \textsf{A} and \textsf{B} in Ref.~\cite{kritsuk..18} and further increasing the resolution would likely eliminate the need to  mask. 

\paragraph{Conclusions.---}
In this Letter, we have shown that MHD simulations of the turbulent, magnetized, multiphase ISM, assuming an optically thin medium at the wavelenghts of interest, perfectly aligned dust grains, and a constant dust-to-gas ratio, lead to predictions for the scale dependence of the $EE$, $BB$, and $TE$ angular power spectra as well as the $E/B$ asymmetry, positive $TE$ correlation and polarization fraction PDF that are consistent with observations by {\em Planck}. Our new high-resolution model demonstrates very tight correspondence of $EE$, $BB$, $TE$ spectra, $E/B$ asymmetry, and polarization fraction with observations. This model, however, does not include parity violating components; hence, we expect to measure zero $TB$ cross-correlation in synthetic polarization maps. This is indeed the case for dust polarization, where $r^{TB}=0.010\pm0.004$ and $-0.004\pm0.004$ for projections perpendicular and parallel to the mean field direction, respectively. It is worth noting that, even at a relatively high grid resolution (i.e., $2112^3$), the $TB$ spectra are extremely noisy and a formal asymptotic standard error of the least-squares fit of time-averaged spectra does not fully reflect the actual statistical uncertainty of the measurement.

While these results validate the conjecture that recent {\em Planck} measurements at 353~GHz reflect the signatures of interstellar turbulence, the origins of the $E/B$ asymmetry and positive $TE$ correlation are still being debated. Results of MHD mode decomposition \cite{beresnyak...13,beresnyak19} for this high-resolution model \cite{beresnyak...24} and for cases \textsf{A} and \textsf{B} from Ref.~\cite{kritsuk..18} indicate dominance of fast magneto-acoustic modes \cite{galtier23} in these supersonic trans-Alfv\'enic models characterized by low plasma beta. From an MHD turbulence theory perspective, Refs.~\cite{caldwell..17,kandel..17,kandel..18} suggest that in such regimes $EE/BB\approx1.9$ and $r_{TE}\approx0.2$, commensurate with predictions from these simulations. Alternative explanations include significant variations of $T$--$E$--$B$ power spectra with column density \cite{bracco...19}, misalignment of filamentary intensity structures \cite{yuen...24} with respect to POS magnetic field \cite{huffenberger..20,clark...21}, and strong nonvanishing skewness of the $E$ polarization \cite{brandenburg......19}.

Finally, the new simulation also allows us to generate detailed self-consistent maps of synchrotron emission, which will be discussed in a separate publication. 

\begin{acknowledgments}
This research was supported in part by the NASA Grant No. 80NSSC22K0724 (A.K. and R.F.) and by the DOE LDRD program under Projects No. LANL 20220107DR (K.W.H.) and No. LANL 20220700PRD1 (K.H.Y.). Computational and storage resources were provided by the ACCESS program (MCA07S014), LRAC allocation at TACC (AST21004), and by the DOE award allocated at NERSC (FES-ERCAP-m4239). This research was also supported in part by Grant No. 216179 from the Simons Foundation and NSF Grant No. PHY-2309135 to the Kavli Institute for Theoretical Physics (KITP).
\end{acknowledgments}

%\bibliography{planck,cmb4}

\begin{thebibliography}{43}%
\makeatletter
\providecommand \@ifxundefined [1]{%
 \@ifx{#1\undefined}
}%
\providecommand \@ifnum [1]{%
 \ifnum #1\expandafter \@firstoftwo
 \else \expandafter \@secondoftwo
 \fi
}%
\providecommand \@ifx [1]{%
 \ifx #1\expandafter \@firstoftwo
 \else \expandafter \@secondoftwo
 \fi
}%
\providecommand \natexlab [1]{#1}%
\providecommand \enquote  [1]{``#1''}%
\providecommand \bibnamefont  [1]{#1}%
\providecommand \bibfnamefont [1]{#1}%
\providecommand \citenamefont [1]{#1}%
\providecommand \href@noop [0]{\@secondoftwo}%
\providecommand \href [0]{\begingroup \@sanitize@url \@href}%
\providecommand \@href[1]{\@@startlink{#1}\@@href}%
\providecommand \@@href[1]{\endgroup#1\@@endlink}%
\providecommand \@sanitize@url [0]{\catcode `\\12\catcode `\$12\catcode
  `\&12\catcode `\#12\catcode `\^12\catcode `\_12\catcode `\%12\relax}%
\providecommand \@@startlink[1]{}%
\providecommand \@@endlink[0]{}%
\providecommand \url  [0]{\begingroup\@sanitize@url \@url }%
\providecommand \@url [1]{\endgroup\@href {#1}{\urlprefix }}%
\providecommand \urlprefix  [0]{URL }%
\providecommand \Eprint [0]{\href }%
\providecommand \doibase [0]{https://doi.org/}%
\providecommand \selectlanguage [0]{\@gobble}%
\providecommand \bibinfo  [0]{\@secondoftwo}%
\providecommand \bibfield  [0]{\@secondoftwo}%
\providecommand \translation [1]{[#1]}%
\providecommand \BibitemOpen [0]{}%
\providecommand \bibitemStop [0]{}%
\providecommand \bibitemNoStop [0]{.\EOS\space}%
\providecommand \EOS [0]{\spacefactor3000\relax}%
\providecommand \BibitemShut  [1]{\csname bibitem#1\endcsname}%
\let\auto@bib@innerbib\@empty
%</preamble>
\bibitem [{\citenamefont {{Kamionkowski}}\ and\ \citenamefont
  {{Kovetz}}(2016)}]{kamionkowski.16}%
  \BibitemOpen
  \bibfield  {author} {\bibinfo {author} {\bibfnamefont {M.}~\bibnamefont
  {{Kamionkowski}}}\ and\ \bibinfo {author} {\bibfnamefont {E.~D.}\
  \bibnamefont {{Kovetz}}},\ }\bibfield  {title} {\bibinfo {title} {{The Quest
  for B Modes from Inflationary Gravitational Waves}},\ }\href
  {https://doi.org/10.1146/annurev-astro-081915-023433} {\bibfield  {journal}
  {\bibinfo  {journal} {Annu. Rev. Astron. Astrophys.}\ }\textbf {\bibinfo
  {volume} {54}},\ \bibinfo {pages} {227} (\bibinfo {year} {2016})},\ \Eprint
  {https://arxiv.org/abs/1510.06042} {arXiv:1510.06042} \BibitemShut {NoStop}%
\bibitem [{\citenamefont {{Dickinson}}(2016)}]{dickinson16}%
  \BibitemOpen
  \bibfield  {author} {\bibinfo {author} {\bibfnamefont {C.}~\bibnamefont
  {{Dickinson}}},\ }\bibfield  {title} {\bibinfo {title} {{CMB foregrounds - A
  brief review}},\ }\href@noop {} {\bibfield  {journal} {\bibinfo  {journal}
  {ArXiv e-prints}\ } (\bibinfo {year} {2016})},\ \Eprint
  {https://arxiv.org/abs/1606.03606} {arXiv:1606.03606} \BibitemShut {NoStop}%
\bibitem [{\citenamefont {{Field}}(1965)}]{field65}%
  \BibitemOpen
  \bibfield  {author} {\bibinfo {author} {\bibfnamefont {G.~B.}\ \bibnamefont
  {{Field}}},\ }\bibfield  {title} {\bibinfo {title} {{Thermal Instability.}},\
  }\href {https://doi.org/10.1086/148317} {\bibfield  {journal} {\bibinfo
  {journal} {\apj}\ }\textbf {\bibinfo {volume} {142}},\ \bibinfo {pages} {531}
  (\bibinfo {year} {1965})}\BibitemShut {NoStop}%
\bibitem [{\citenamefont {{Goldsmith}}\ \emph {et~al.}(1969)\citenamefont
  {{Goldsmith}}, \citenamefont {{Habing}},\ and\ \citenamefont
  {{Field}}}]{goldsmith..69}%
  \BibitemOpen
  \bibfield  {author} {\bibinfo {author} {\bibfnamefont {D.~W.}\ \bibnamefont
  {{Goldsmith}}}, \bibinfo {author} {\bibfnamefont {H.~J.}\ \bibnamefont
  {{Habing}}},\ and\ \bibinfo {author} {\bibfnamefont {G.~B.}\ \bibnamefont
  {{Field}}},\ }\bibfield  {title} {\bibinfo {title} {{Thermal Properties of
  Interstellar Gas Heated by Cosmic Rays}},\ }\href
  {https://doi.org/10.1086/150181} {\bibfield  {journal} {\bibinfo  {journal}
  {\apj}\ }\textbf {\bibinfo {volume} {158}},\ \bibinfo {pages} {173} (\bibinfo
  {year} {1969})}\BibitemShut {NoStop}%
\bibitem [{\citenamefont {{Zel'Dovich}}\ and\ \citenamefont
  {{Pikel'Ner}}(1969)}]{zeldovich.69}%
  \BibitemOpen
  \bibfield  {author} {\bibinfo {author} {\bibfnamefont {Y.~B.}\ \bibnamefont
  {{Zel'Dovich}}}\ and\ \bibinfo {author} {\bibfnamefont {S.~B.}\ \bibnamefont
  {{Pikel'Ner}}},\ }\bibfield  {title} {\bibinfo {title} {{The Phase
  Equilibrium and Dynamics of a Gas Volume That is Heated and Cooled}},\
  }\href@noop {} {\bibfield  {journal} {\bibinfo  {journal} {Sov. J.
  Exp. Theor. Phys.}\ }\textbf {\bibinfo {volume} {29}},\
  \bibinfo {pages} {170} (\bibinfo {year} {1969})}\BibitemShut {NoStop}%
\bibitem [{\citenamefont {{Wolfire}}\ \emph {et~al.}(2003)\citenamefont
  {{Wolfire}}, \citenamefont {{McKee}}, \citenamefont {{Hollenbach}},\ and\
  \citenamefont {{Tielens}}}]{wolfire...03}%
  \BibitemOpen
  \bibfield  {author} {\bibinfo {author} {\bibfnamefont {M.~G.}\ \bibnamefont
  {{Wolfire}}}, \bibinfo {author} {\bibfnamefont {C.~F.}\ \bibnamefont
  {{McKee}}}, \bibinfo {author} {\bibfnamefont {D.}~\bibnamefont
  {{Hollenbach}}},\ and\ \bibinfo {author} {\bibfnamefont {A.~G.~G.~M.}\
  \bibnamefont {{Tielens}}},\ }\bibfield  {title} {\bibinfo {title} {{Neutral
  Atomic Phases of the Interstellar Medium in the Galaxy}},\ }\href
  {https://doi.org/10.1086/368016} {\bibfield  {journal} {\bibinfo  {journal}
  {\apj}\ }\textbf {\bibinfo {volume} {587}},\ \bibinfo {pages} {278} (\bibinfo
  {year} {2003})},\ \Eprint {https://arxiv.org/abs/astro-ph/0207098}
  {arXiv:astro-ph/0207098} \BibitemShut {NoStop}%
\bibitem [{\citenamefont {{Mac Low}}\ and\ \citenamefont
  {{Klessen}}(2004)}]{maclow.04}%
  \BibitemOpen
  \bibfield  {author} {\bibinfo {author} {\bibfnamefont {M.}~\bibnamefont {{Mac
  Low}}}\ and\ \bibinfo {author} {\bibfnamefont {R.~S.}\ \bibnamefont
  {{Klessen}}},\ }\bibfield  {title} {\bibinfo {title} {{Control of star
  formation by supersonic turbulence}},\ }\href@noop {} {\bibfield  {journal}
  {\bibinfo  {journal} {Rev. Mod. Phys.}\ }\textbf {\bibinfo {volume} {76}},\
  \bibinfo {pages} {125} (\bibinfo {year} {2004})}\BibitemShut {NoStop}%
\bibitem [{\citenamefont {{Hennebelle}}\ and\ \citenamefont
  {{Falgarone}}(2012)}]{hennebelle.12}%
  \BibitemOpen
  \bibfield  {author} {\bibinfo {author} {\bibfnamefont {P.}~\bibnamefont
  {{Hennebelle}}}\ and\ \bibinfo {author} {\bibfnamefont {E.}~\bibnamefont
  {{Falgarone}}},\ }\bibfield  {title} {\bibinfo {title} {{Turbulent molecular
  clouds}},\ }\href {https://doi.org/10.1007/s00159-012-0055-y} {\bibfield
  {journal} {\bibinfo  {journal} {\aapr}\ }\textbf {\bibinfo {volume} {20}},\
  \bibinfo {eid} {55} (\bibinfo {year} {2012})},\ \Eprint
  {https://arxiv.org/abs/1211.0637} {arXiv:1211.0637} \BibitemShut {NoStop}%
\bibitem [{\citenamefont {Klessen}\ and\ \citenamefont
  {Glover}(2016)}]{klessen.16}%
  \BibitemOpen
  \bibfield  {author} {\bibinfo {author} {\bibfnamefont {R.~S.}\ \bibnamefont
  {Klessen}}\ and\ \bibinfo {author} {\bibfnamefont {S.~C.~O.}\ \bibnamefont
  {Glover}},\ }\bibinfo {title} {Physical processes in the interstellar
  medium},\ in\ \href {https://doi.org/10.1007/978-3-662-47890-5_2} {\emph
  {\bibinfo {booktitle} {Star Formation in Galaxy Evolution: Connecting
  Numerical Models to Reality: Saas-Fee Advanced Course 43. Swiss Society for
  Astrophysics and Astronomy}}},\ \bibinfo {editor} {edited by\ \bibinfo
  {editor} {\bibfnamefont {Y.}~\bibnamefont {Revaz}}, \bibinfo {editor}
  {\bibfnamefont {P.}~\bibnamefont {Jablonka}}, \bibinfo {editor}
  {\bibfnamefont {R.}~\bibnamefont {Teyssier}},\ and\ \bibinfo {editor}
  {\bibfnamefont {L.}~\bibnamefont {Mayer}}}\ (\bibinfo  {publisher} {Springer
  Berlin Heidelberg},\ \bibinfo {address} {Berlin, Heidelberg},\ \bibinfo
  {year} {2016})\ pp.\ \bibinfo {pages} {85--249}\BibitemShut {NoStop}%
\bibitem [{\citenamefont {{Colman}}\ \emph {et~al.}(2022)\citenamefont
  {{Colman}}, \citenamefont {{Robitaille}}, \citenamefont {{Hennebelle}},
  \citenamefont {{Miville-Desch{\^e}nes}}, \citenamefont {{Brucy}},
  \citenamefont {{Klessen}}, \citenamefont {{Glover}}, \citenamefont {{Soler}},
  \citenamefont {{Elia}}, \citenamefont {{Traficante}}, \citenamefont
  {{Molinari}},\ and\ \citenamefont {{Testi}}}]{colman.........22}%
  \BibitemOpen
  \bibfield  {author} {\bibinfo {author} {\bibfnamefont {T.}~\bibnamefont
  {{Colman}}}, \bibinfo {author} {\bibfnamefont {J.-F.}\ \bibnamefont
  {{Robitaille}}}, \bibinfo {author} {\bibfnamefont {P.}~\bibnamefont
  {{Hennebelle}}}, \bibinfo {author} {\bibfnamefont {M.-A.}\ \bibnamefont
  {{Miville-Desch{\^e}nes}}}, \bibinfo {author} {\bibfnamefont
  {N.}~\bibnamefont {{Brucy}}}, \bibinfo {author} {\bibfnamefont {R.~S.}\
  \bibnamefont {{Klessen}}}, \bibinfo {author} {\bibfnamefont {S.~C.~O.}\
  \bibnamefont {{Glover}}}, \bibinfo {author} {\bibfnamefont {J.~D.}\
  \bibnamefont {{Soler}}}, \bibinfo {author} {\bibfnamefont {D.}~\bibnamefont
  {{Elia}}}, \bibinfo {author} {\bibfnamefont {A.}~\bibnamefont
  {{Traficante}}}, \bibinfo {author} {\bibfnamefont {S.}~\bibnamefont
  {{Molinari}}},\ and\ \bibinfo {author} {\bibfnamefont {L.}~\bibnamefont
  {{Testi}}},\ }\bibfield  {title} {\bibinfo {title} {{The signature of
  large-scale turbulence driving on the structure of the interstellar
  medium}},\ }\href {https://doi.org/10.1093/mnras/stac1543} {\bibfield
  {journal} {\bibinfo  {journal} {\mnras}\ }\textbf {\bibinfo {volume} {514}},\
  \bibinfo {pages} {3670} (\bibinfo {year} {2022})},\ \Eprint
  {https://arxiv.org/abs/2206.00451} {arXiv:2206.00451}
  \BibitemShut {NoStop}%
\bibitem [{\citenamefont {{Elmegreen}}(2024)}]{elmegreen24}%
  \BibitemOpen
  \bibfield  {author} {\bibinfo {author} {\bibfnamefont {B.~G.}\ \bibnamefont
  {{Elmegreen}}},\ }\bibfield  {title} {\bibinfo {title} {{Feedback and Galaxy
  Dynamics: A Study of Turbulence and Star Formation in 34 Galaxies Using the
  PHANGS Survey}},\ }\href {https://doi.org/10.3847/1538-4357/ad36c7}
  {\bibfield  {journal} {\bibinfo  {journal} {\apj}\ }\textbf {\bibinfo
  {volume} {966}},\ \bibinfo {eid} {233} (\bibinfo {year} {2024})},\ \Eprint
  {https://arxiv.org/abs/2403.12927} {arXiv:2403.12927}
  \BibitemShut {NoStop}%
\bibitem [{\citenamefont {{Kritsuk}}\ \emph {et~al.}(2017)\citenamefont
  {{Kritsuk}}, \citenamefont {{Ustyugov}},\ and\ \citenamefont
  {{Norman}}}]{kritsuk..17}%
  \BibitemOpen
  \bibfield  {author} {\bibinfo {author} {\bibfnamefont {A.~G.}\ \bibnamefont
  {{Kritsuk}}}, \bibinfo {author} {\bibfnamefont {S.~D.}\ \bibnamefont
  {{Ustyugov}}},\ and\ \bibinfo {author} {\bibfnamefont {M.~L.}\ \bibnamefont
  {{Norman}}},\ }\bibfield  {title} {\bibinfo {title} {{The structure and
  statistics of interstellar turbulence}},\ }\href
  {https://doi.org/10.1088/1367-2630/aa7156} {\bibfield  {journal} {\bibinfo
  {journal} {N. J. Phys.}\ }\textbf {\bibinfo {volume} {19}},\ \bibinfo {eid}
  {065003} (\bibinfo {year} {2017})},\ \Eprint
  {https://arxiv.org/abs/1705.01912} {arXiv:1705.01912} \BibitemShut {NoStop}%
\bibitem [{\citenamefont {{Kritsuk}}\ \emph {et~al.}(2018)\citenamefont
  {{Kritsuk}}, \citenamefont {{Flauger}},\ and\ \citenamefont
  {{Ustyugov}}}]{kritsuk..18}%
  \BibitemOpen
  \bibfield  {author} {\bibinfo {author} {\bibfnamefont {A.~G.}\ \bibnamefont
  {{Kritsuk}}}, \bibinfo {author} {\bibfnamefont {R.}~\bibnamefont
  {{Flauger}}},\ and\ \bibinfo {author} {\bibfnamefont {S.~D.}\ \bibnamefont
  {{Ustyugov}}},\ }\bibfield  {title} {\bibinfo {title} {{Dust-Polarization
  Maps for Local Interstellar Turbulence}},\ }\href
  {https://doi.org/10.1103/PhysRevLett.121.021104} {\bibfield  {journal}
  {\bibinfo  {journal} {\prl}\ }\textbf {\bibinfo {volume} {121}},\ \bibinfo
  {eid} {021104} (\bibinfo {year} {2018})},\ \Eprint
  {https://arxiv.org/abs/1711.11108} {arXiv:1711.11108}
  \BibitemShut {NoStop}%
\bibitem [{\citenamefont {{Planck Collaboration XI}}(2020)}]{pr3.20}%
  \BibitemOpen
  \bibfield  {author} {\bibinfo {author} {\bibnamefont {{Planck Collaboration
  XI}}},\ }\bibfield  {title} {\bibinfo {title} {{Planck 2018 results. XI.
  Polarized dust foregrounds}},\ }\href
  {https://doi.org/10.1051/0004-6361/201832618} {\bibfield  {journal} {\bibinfo
   {journal} {\aap}\ }\textbf {\bibinfo {volume} {641}},\ \bibinfo {eid} {A11}
  (\bibinfo {year} {2020})},\ \Eprint {https://arxiv.org/abs/1801.04945}
  {arXiv:1801.04945} \BibitemShut {NoStop}%
\bibitem [{\citenamefont {{Kritsuk}}\ \emph {et~al.}(2010)\citenamefont
  {{Kritsuk}}, \citenamefont {{Ustyugov}}, \citenamefont {{Norman}},\ and\
  \citenamefont {{Padoan}}}]{kritsuk...10}%
  \BibitemOpen
  \bibfield  {author} {\bibinfo {author} {\bibfnamefont {A.~G.}\ \bibnamefont
  {{Kritsuk}}}, \bibinfo {author} {\bibfnamefont {S.~D.}\ \bibnamefont
  {{Ustyugov}}}, \bibinfo {author} {\bibfnamefont {M.~L.}\ \bibnamefont
  {{Norman}}},\ and\ \bibinfo {author} {\bibfnamefont {P.}~\bibnamefont
  {{Padoan}}},\ }\bibfield  {title} {\bibinfo {title} {{Self-organization in
  Turbulent Molecular Clouds: Compressional Versus Solenoidal Modes}},\ }in\
  \href {https://doi.org/10.48550/arXiv.0912.0546} {\emph {\bibinfo {booktitle}
  {Numerical Modeling of Space Plasma Flows, Astronum-2009}}},\ \bibinfo
  {series} {Astronomical Society of the Pacific Conference Series}, Vol.\
  \bibinfo {volume} {429},\ \bibinfo {editor} {edited by\ \bibinfo {editor}
  {\bibfnamefont {N.~V.}\ \bibnamefont {{Pogorelov}}}, \bibinfo {editor}
  {\bibfnamefont {E.}~\bibnamefont {{Audit}}},\ and\ \bibinfo {editor}
  {\bibfnamefont {G.~P.}\ \bibnamefont {{Zank}}}}\ (\bibinfo {year} {2010})\
  p.~\bibinfo {pages} {15},\ \Eprint {https://arxiv.org/abs/0912.0546}
  {arXiv:0912.0546} \BibitemShut {NoStop}%
\bibitem [{Note1()}]{Note1}%
  \BibitemOpen
  \bibinfo {note} {Note the new simulation employs methods with
  piecewise-linear (PLM, \cite {mignone14}) reconstruction, while previous
  lower resolution models used piecewise-parabolic reconstruction (PPML, \cite
  {ustyugov...09}). {\protect \leavevmode {\protect \color {black}\protect
  Both codes used the HLLD Riemann solver.}}}\BibitemShut {Stop}%
\bibitem [{\citenamefont {{Sytine}}\ \emph {et~al.}(2000)\citenamefont
  {{Sytine}}, \citenamefont {{Porter}}, \citenamefont {{Woodward}},
  \citenamefont {{Hodson}},\ and\ \citenamefont {{Winkler}}}]{sytine....00}%
  \BibitemOpen
  \bibfield  {author} {\bibinfo {author} {\bibfnamefont {I.~V.}\ \bibnamefont
  {{Sytine}}}, \bibinfo {author} {\bibfnamefont {D.~H.}\ \bibnamefont
  {{Porter}}}, \bibinfo {author} {\bibfnamefont {P.~R.}\ \bibnamefont
  {{Woodward}}}, \bibinfo {author} {\bibfnamefont {S.~W.}\ \bibnamefont
  {{Hodson}}},\ and\ \bibinfo {author} {\bibfnamefont {K.-H.}\ \bibnamefont
  {{Winkler}}},\ }\bibfield  {title} {\bibinfo {title} {{Convergence Tests for
  the Piecewise Parabolic Method and Navier-Stokes Solutions for Homogeneous
  Compressible Turbulence}},\ }\href {https://doi.org/10.1006/jcph.1999.6416}
  {\bibfield  {journal} {\bibinfo  {journal} {\jcph}\ }\textbf {\bibinfo
  {volume} {158}},\ \bibinfo {pages} {225} (\bibinfo {year}
  {2000})}\BibitemShut {NoStop}%
\bibitem [{Note2()}]{Note2}%
  \BibitemOpen
  \bibinfo {note} {E.g., the bottleneck phenomenon \cite
  {falkovich94}}\BibitemShut {NoStop}%
\bibitem [{\citenamefont {Cabral}\ and\ \citenamefont
  {Leedom}(1993)}]{cabral.93}%
  \BibitemOpen
  \bibfield  {author} {\bibinfo {author} {\bibfnamefont {B.}~\bibnamefont
  {Cabral}}\ and\ \bibinfo {author} {\bibfnamefont {L.~C.}\ \bibnamefont
  {Leedom}},\ }\bibfield  {title} {\bibinfo {title} {{Imaging vector fields
  using line integral convolution}},\ }\href@noop {} {\bibfield  {journal}
  {\bibinfo  {journal} {{Proc. 20th Annual Conference on Computer Graphics and
  Interactive Techniques}}\ ,\ \bibinfo {pages} {263}} (\bibinfo {year}
  {1993})}\BibitemShut {NoStop}%
\bibitem [{\citenamefont {{Kritsuk}}\ \emph {et~al.}(2011)\citenamefont
  {{Kritsuk}}, \citenamefont {{Ustyugov}},\ and\ \citenamefont
  {{Norman}}}]{kritsuk..11}%
  \BibitemOpen
  \bibfield  {author} {\bibinfo {author} {\bibfnamefont {A.~G.}\ \bibnamefont
  {{Kritsuk}}}, \bibinfo {author} {\bibfnamefont {S.~D.}\ \bibnamefont
  {{Ustyugov}}},\ and\ \bibinfo {author} {\bibfnamefont {M.~L.}\ \bibnamefont
  {{Norman}}},\ }\bibfield  {title} {\bibinfo {title} {{Interstellar Turbulence
  and Star Formation}},\ }in\ \href {https://doi.org/10.1017/S1743921311000342}
  {\emph {\bibinfo {booktitle} {Computational Star Formation}}},\ Vol.\
  \bibinfo {volume} {270},\ \bibinfo {editor} {edited by\ \bibinfo {editor}
  {\bibfnamefont {J.}~\bibnamefont {{Alves}}}, \bibinfo {editor} {\bibfnamefont
  {B.~G.}\ \bibnamefont {{Elmegreen}}}, \bibinfo {editor} {\bibfnamefont
  {J.~M.}\ \bibnamefont {{Girart}}},\ and\ \bibinfo {editor} {\bibfnamefont
  {V.}~\bibnamefont {{Trimble}}}}\ (\bibinfo {year} {2011})\ pp.\ \bibinfo
  {pages} {179--186},\ \Eprint {https://arxiv.org/abs/1011.2177}
  {arXiv:1011.2177} \BibitemShut {NoStop}%
\bibitem [{\citenamefont {{Cho}}(2023)}]{cho.23}%
  \BibitemOpen
  \bibfield  {author} {\bibinfo {author} {\bibfnamefont {J.}~\bibnamefont
  {{Cho}}},\ }\bibfield  {title} {\bibinfo {title} {{Obtaining the Strength of
  the Magnetic Field from $E$- and $B$-Modes of Dust Polarization}},\ }\href
  {https://doi.org/10.3847/1538-4357/ace10a} {\bibfield  {journal} {\bibinfo
  {journal} {\apj}\ }\textbf {\bibinfo {volume} {953}},\ \bibinfo {eid} {114}
  (\bibinfo {year} {2023})},\ \Eprint {https://arxiv.org/abs/2306.11571}
  {arXiv:2306.11571} \BibitemShut {NoStop}%
\bibitem [{\citenamefont {{Stalpes}}\ \emph {et~al.}(2024)\citenamefont
  {{Stalpes}}, \citenamefont {{Collins}},\ and\ \citenamefont
  {{Huffenberger}}}]{stalpes..24}%
  \BibitemOpen
  \bibfield  {author} {\bibinfo {author} {\bibfnamefont {K.~A.}\ \bibnamefont
  {{Stalpes}}}, \bibinfo {author} {\bibfnamefont {D.~C.}\ \bibnamefont
  {{Collins}}},\ and\ \bibinfo {author} {\bibfnamefont {K.~M.}\ \bibnamefont
  {{Huffenberger}}},\ }\bibfield  {title} {\bibinfo {title} {{Planck Dust
  Polarization Power Spectra Are Consistent with Strongly Supersonic
  Turbulence}},\ }\href {https://doi.org/10.3847/1538-4357/ad571b} {\bibfield
  {journal} {\bibinfo  {journal} {\apj}\ }\textbf {\bibinfo {volume} {972}},\
  \bibinfo {eid} {26} (\bibinfo {year} {2024})},\ \Eprint
  {https://arxiv.org/abs/2404.02874} {arXiv:2404.02874}
  \BibitemShut {NoStop}%
\bibitem [{Note3()}]{Note3}%
  \BibitemOpen
  \bibinfo {note} {{\protect \leavevmode {\protect \color {black}\protect To
  accurately measure statistics of polarized foregrounds in our model for
  stationary and homogeneous ISM turbulence, we use ensemble averages for a
  reasonably large set of high-resolution synthetic maps sampled over a period
  of $9\tau _{\protect \rm d}=45$~Myr. We then rely on the ergodic hypothesis
  (see, e.g., Ref.~\cite {monin.65}) to justify the comparison with observational
  data for LR71---the largest region in the {\protect \em Planck} sky, for
  which accuracy is achieved through extensive space averaging}}}\BibitemShut
{NoStop}%
\bibitem [{Note4()}]{Note4}%
  \BibitemOpen
  \bibinfo {note} {\textcolor{black}{It is worth noting that formal standard deviations provided
  here and below for the model-based spectral indices (and spectral ratios)
  only reflect the quality of the least-squares fit of the average spectrum for
  the projection along the mean field in the specified interval of wave
  numbers. The accuracy of these spectral indices depends on: (i) the chosen
  interval of wave numbers, which we deem as inertial range; (ii) the unknown
  orientation of the mean magnetic field with respect to the line of sight; (iii) the
  variation of slope between individual flow snapshots; (iv) the number of
  correlation times in the time-averaging interval; etc. Therefore, it is not
  hard to imagine that the actual standard deviation of our measurement is of
  the order of 0.05 or larger.}}\BibitemShut {Stop}%
\bibitem [{\citenamefont {{Planck Collaboration Int.
  XIX}}(2015)}]{planckXIX.15}%
  \BibitemOpen
  \bibfield  {author} {\bibinfo {author} {\bibnamefont {{Planck Collaboration
  Int. XIX}}},\ }\bibfield  {title} {\bibinfo {title} {{Planck intermediate
  results. XIX. An overview of the polarized thermal emission from Galactic
  dust}},\ }\href {https://doi.org/10.1051/0004-6361/201424082} {\bibfield
  {journal} {\bibinfo  {journal} {\aap}\ }\textbf {\bibinfo {volume} {576}},\
  \bibinfo {eid} {A104} (\bibinfo {year} {2015})},\ \Eprint
  {https://arxiv.org/abs/1405.0871} {arXiv:1405.0871}
  \BibitemShut {NoStop}%
\bibitem [{\citenamefont {{Planck Collaboration Int.
  XLIV}}(2016)}]{planckXLIV.16}%
  \BibitemOpen
  \bibfield  {author} {\bibinfo {author} {\bibnamefont {{Planck Collaboration
  Int. XLIV}}},\ }\bibfield  {title} {\bibinfo {title} {{Planck intermediate
  results. XLIV. Structure of the Galactic magnetic field from dust
  polarization maps of the southern Galactic cap}},\ }\href
  {https://doi.org/10.1051/0004-6361/201628636} {\bibfield  {journal} {\bibinfo
   {journal} {\aap}\ }\textbf {\bibinfo {volume} {596}},\ \bibinfo {eid} {A105}
  (\bibinfo {year} {2016})},\ \Eprint {https://arxiv.org/abs/1604.01029}
  {arXiv:1604.01029} \BibitemShut {NoStop}%
\bibitem [{\citenamefont {{Planck Collaboration XII}}(2020)}]{planckXII.20}%
  \BibitemOpen
  \bibfield  {author} {\bibinfo {author} {\bibnamefont {{Planck Collaboration
  XII}}},\ }\bibfield  {title} {\bibinfo {title} {{Planck 2018 results. XII.
  Galactic astrophysics using polarized dust emission}},\ }\href
  {https://doi.org/10.1051/0004-6361/201833885} {\bibfield  {journal} {\bibinfo
   {journal} {\aap}\ }\textbf {\bibinfo {volume} {641}},\ \bibinfo {eid} {A12}
  (\bibinfo {year} {2020})},\ \Eprint {https://arxiv.org/abs/1807.06212}
  {arXiv:1807.06212} \BibitemShut {NoStop}%
\bibitem [{\citenamefont {{Planck Collaboration Int.
  XXX}}(2016)}]{planckXXX.16}%
  \BibitemOpen
  \bibfield  {author} {\bibinfo {author} {\bibnamefont {{Planck Collaboration
  Int. XXX}}},\ }\bibfield  {title} {\bibinfo {title} {{Planck intermediate
  results. XXX. The angular power spectrum of polarized dust emission at
  intermediate and high Galactic latitudes}},\ }\href
  {https://doi.org/10.1051/0004-6361/201425034} {\bibfield  {journal} {\bibinfo
   {journal} {\aap}\ }\textbf {\bibinfo {volume} {586}},\ \bibinfo {eid} {A133}
  (\bibinfo {year} {2016})},\ \Eprint {https://arxiv.org/abs/1409.5738}
  {arXiv:1409.5738} \BibitemShut {NoStop}%
\bibitem [{\citenamefont {{Beresnyak}}\ \emph {et~al.}(2013)\citenamefont
  {{Beresnyak}}, \citenamefont {{Xu}}, \citenamefont {{Li}},\ and\
  \citenamefont {{Schlickeiser}}}]{beresnyak...13}%
  \BibitemOpen
  \bibfield  {author} {\bibinfo {author} {\bibfnamefont {A.}~\bibnamefont
  {{Beresnyak}}}, \bibinfo {author} {\bibfnamefont {H.}~\bibnamefont {{Xu}}},
  \bibinfo {author} {\bibfnamefont {H.}~\bibnamefont {{Li}}},\ and\ \bibinfo
  {author} {\bibfnamefont {R.}~\bibnamefont {{Schlickeiser}}},\ }\bibfield
  {title} {\bibinfo {title} {{Magnetohydrodynamic Turbulence and Cosmic-Ray
  Reacceleration in Galaxy Clusters}},\ }\href
  {https://doi.org/10.1088/0004-637X/771/2/131} {\bibfield  {journal} {\bibinfo
   {journal} {\apj}\ }\textbf {\bibinfo {volume} {771}},\ \bibinfo {eid} {131}
  (\bibinfo {year} {2013})},\ \Eprint {https://arxiv.org/abs/1301.7453}
  {arXiv:1301.7453} \BibitemShut {NoStop}%
\bibitem [{\citenamefont {{Beresnyak}}(2019)}]{beresnyak19}%
  \BibitemOpen
  \bibfield  {author} {\bibinfo {author} {\bibfnamefont {A.}~\bibnamefont
  {{Beresnyak}}},\ }\bibfield  {title} {\bibinfo {title} {{MHD turbulence}},\
  }\href {https://doi.org/10.1007/s41115-019-0005-8} {\bibfield  {journal}
  {\bibinfo  {journal} {Living Rev. Comput. Astrophys.}\ }\textbf
  {\bibinfo {volume} {5}},\ \bibinfo {eid} {2} (\bibinfo {year} {2019})},\
  \Eprint {https://arxiv.org/abs/1910.03585} {arXiv:1910.03585}
  \BibitemShut {NoStop}%
\bibitem [{\citenamefont {{Beresnyak}}\ \emph {et~al.}(2024)\citenamefont
  {{Beresnyak}}, \citenamefont {{Kritsuk}}, \citenamefont {{Yuen}},\ and\
  \citenamefont {{Ho}}}]{beresnyak...24}%
  \BibitemOpen
  \bibfield  {author} {\bibinfo {author} {\bibfnamefont {A.}~\bibnamefont
  {{Beresnyak}}}, \bibinfo {author} {\bibfnamefont {A.~G.}\ \bibnamefont
  {{Kritsuk}}}, \bibinfo {author} {\bibfnamefont {K.~H.}\ \bibnamefont
  {{Yuen}}},\ and\ \bibinfo {author} {\bibfnamefont {K.~W.}\ \bibnamefont
  {{Ho}}} }\href@noop {}\ \bibinfo {note} {(to be published).}\BibitemShut {Stop}%
\bibitem [{\citenamefont {{Galtier}}(2023)}]{galtier23}%
  \BibitemOpen
  \bibfield  {author} {\bibinfo {author} {\bibfnamefont {S.}~\bibnamefont
  {{Galtier}}},\ }\bibfield  {title} {\bibinfo {title} {{Fast magneto-acoustic
  wave turbulence and the Iroshnikov-Kraichnan spectrum}},\ }\href
  {https://doi.org/10.1017/S0022377823000259} {\bibfield  {journal} {\bibinfo
  {journal} {J. Plasma Phys.}\ }\textbf {\bibinfo {volume} {89}},\ \bibinfo
  {eid} {905890205} (\bibinfo {year} {2023})},\ \Eprint
  {https://arxiv.org/abs/2303.00643} {arXiv:2303.00643}
  \BibitemShut {NoStop}%
\bibitem [{\citenamefont {{Caldwell}}\ \emph {et~al.}(2017)\citenamefont
  {{Caldwell}}, \citenamefont {{Hirata}},\ and\ \citenamefont
  {{Kamionkowski}}}]{caldwell..17}%
  \BibitemOpen
  \bibfield  {author} {\bibinfo {author} {\bibfnamefont {R.~R.}\ \bibnamefont
  {{Caldwell}}}, \bibinfo {author} {\bibfnamefont {C.}~\bibnamefont
  {{Hirata}}},\ and\ \bibinfo {author} {\bibfnamefont {M.}~\bibnamefont
  {{Kamionkowski}}},\ }\bibfield  {title} {\bibinfo {title} {{Dust-polarization
  Maps and Interstellar Turbulence}},\ }\href
  {https://doi.org/10.3847/1538-4357/aa679c} {\bibfield  {journal} {\bibinfo
  {journal} {\apj}\ }\textbf {\bibinfo {volume} {839}},\ \bibinfo {eid} {91}
  (\bibinfo {year} {2017})},\ \Eprint {https://arxiv.org/abs/1608.08138}
  {arXiv:1608.08138} \BibitemShut {NoStop}%
\bibitem [{\citenamefont {{Kandel}}\ \emph {et~al.}(2017)\citenamefont
  {{Kandel}}, \citenamefont {{Lazarian}},\ and\ \citenamefont
  {{Pogosyan}}}]{kandel..17}%
  \BibitemOpen
  \bibfield  {author} {\bibinfo {author} {\bibfnamefont {D.}~\bibnamefont
  {{Kandel}}}, \bibinfo {author} {\bibfnamefont {A.}~\bibnamefont
  {{Lazarian}}},\ and\ \bibinfo {author} {\bibfnamefont {D.}~\bibnamefont
  {{Pogosyan}}},\ }\bibfield  {title} {\bibinfo {title} {{Can the observed E/B
  ratio for dust galactic foreground be explained by sub-Alfv{\'e}nic
  turbulence?}},\ }\href {https://doi.org/10.1093/mnrasl/slx128} {\bibfield
  {journal} {\bibinfo  {journal} {\mnras}\ }\textbf {\bibinfo {volume} {472}},\
  \bibinfo {pages} {L10} (\bibinfo {year} {2017})},\ \Eprint
  {https://arxiv.org/abs/1707.06276} {arXiv:1707.06276} \BibitemShut {NoStop}%
\bibitem [{\citenamefont {{Kandel}}\ \emph {et~al.}(2018)\citenamefont
  {{Kandel}}, \citenamefont {{Lazarian}},\ and\ \citenamefont
  {{Pogosyan}}}]{kandel..18}%
  \BibitemOpen
  \bibfield  {author} {\bibinfo {author} {\bibfnamefont {D.}~\bibnamefont
  {{Kandel}}}, \bibinfo {author} {\bibfnamefont {A.}~\bibnamefont
  {{Lazarian}}},\ and\ \bibinfo {author} {\bibfnamefont {D.}~\bibnamefont
  {{Pogosyan}}},\ }\bibfield  {title} {\bibinfo {title} {{Statistical
  properties of Galactic CMB foregrounds: Dust and synchrotron}},\ }\href
  {https://doi.org/10.1093/mnras/sty1115} {\bibfield  {journal} {\bibinfo
  {journal} {\mnras}\ }\textbf {\bibinfo {volume} {478}},\ \bibinfo {pages}
  {530} (\bibinfo {year} {2018})},\ \Eprint {https://arxiv.org/abs/1711.03161}
  {arXiv:1711.03161} \BibitemShut {NoStop}%
\bibitem [{\citenamefont {{Bracco}}\ \emph {et~al.}(2019)\citenamefont
  {{Bracco}}, \citenamefont {{Ghosh}}, \citenamefont {{Boulanger}},\ and\
  \citenamefont {{Aumont}}}]{bracco...19}%
  \BibitemOpen
  \bibfield  {author} {\bibinfo {author} {\bibfnamefont {A.}~\bibnamefont
  {{Bracco}}}, \bibinfo {author} {\bibfnamefont {T.}~\bibnamefont {{Ghosh}}},
  \bibinfo {author} {\bibfnamefont {F.}~\bibnamefont {{Boulanger}}},\ and\
  \bibinfo {author} {\bibfnamefont {J.}~\bibnamefont {{Aumont}}},\ }\bibfield
  {title} {\bibinfo {title} {{Link between $E-B$ polarization modes and gas
  column density from interstellar dust emission}},\ }\href
  {https://doi.org/10.1051/0004-6361/201935951} {\bibfield  {journal} {\bibinfo
   {journal} {\aap}\ }\textbf {\bibinfo {volume} {632}},\ \bibinfo {eid} {A17}
  (\bibinfo {year} {2019})},\ \Eprint {https://arxiv.org/abs/1905.10471}
  {arXiv:1905.10471} \BibitemShut {NoStop}%
\bibitem [{\citenamefont {{Yuen}}\ \emph {et~al.}(2024)\citenamefont {{Yuen}},
  \citenamefont {{Ho}}, \citenamefont {{Law}},\ and\ \citenamefont
  {{Chen}}}]{yuen...24}%
  \BibitemOpen
  \bibfield  {author} {\bibinfo {author} {\bibfnamefont {K.~H.}\ \bibnamefont
  {{Yuen}}}, \bibinfo {author} {\bibfnamefont {K.~W.}\ \bibnamefont {{Ho}}},
  \bibinfo {author} {\bibfnamefont {C.~Y.}\ \bibnamefont {{Law}}},\ and\
  \bibinfo {author} {\bibfnamefont {A.}~\bibnamefont {{Chen}}},\ }\bibfield
  {title} {\bibinfo {title} {{Neutral hydrogen filaments in interstellar media:
  Are they physical?}},\ }\href {https://doi.org/10.1007/s41614-024-00156-5}
  {\bibfield  {journal} {\bibinfo  {journal} {Rev. Mod. Plasma Phys.}\ }\textbf
  {\bibinfo {volume} {8}},\ \bibinfo {eid} {21} (\bibinfo {year} {2024})},\
  \Eprint {https://arxiv.org/abs/2404.19101} {arXiv:2404.19101}
  \BibitemShut {NoStop}%
\bibitem [{\citenamefont {{Huffenberger}}\ \emph {et~al.}(2020)\citenamefont
  {{Huffenberger}}, \citenamefont {{Rotti}},\ and\ \citenamefont
  {{Collins}}}]{huffenberger..20}%
  \BibitemOpen
  \bibfield  {author} {\bibinfo {author} {\bibfnamefont {K.~M.}\ \bibnamefont
  {{Huffenberger}}}, \bibinfo {author} {\bibfnamefont {A.}~\bibnamefont
  {{Rotti}}},\ and\ \bibinfo {author} {\bibfnamefont {D.~C.}\ \bibnamefont
  {{Collins}}},\ }\bibfield  {title} {\bibinfo {title} {{The Power Spectra of
  Polarized, Dusty Filaments}},\ }\href
  {https://doi.org/10.3847/1538-4357/ab9df9} {\bibfield  {journal} {\bibinfo
  {journal} {\apj}\ }\textbf {\bibinfo {volume} {899}},\ \bibinfo {eid} {31}
  (\bibinfo {year} {2020})},\ \Eprint {https://arxiv.org/abs/1906.10052}
  {arXiv:1906.10052} \BibitemShut {NoStop}%
\bibitem [{\citenamefont {{Clark}}\ \emph {et~al.}(2021)\citenamefont
  {{Clark}}, \citenamefont {{Kim}}, \citenamefont {{Hill}},\ and\ \citenamefont
  {{Hensley}}}]{clark...21}%
  \BibitemOpen
  \bibfield  {author} {\bibinfo {author} {\bibfnamefont {S.~E.}\ \bibnamefont
  {{Clark}}}, \bibinfo {author} {\bibfnamefont {C.-G.}\ \bibnamefont {{Kim}}},
  \bibinfo {author} {\bibfnamefont {J.~C.}\ \bibnamefont {{Hill}}},\ and\
  \bibinfo {author} {\bibfnamefont {B.~S.}\ \bibnamefont {{Hensley}}},\
  }\bibfield  {title} {\bibinfo {title} {{The Origin of Parity Violation in
  Polarized Dust Emission and Implications for Cosmic Birefringence}},\ }\href
  {https://doi.org/10.3847/1538-4357/ac0e35} {\bibfield  {journal} {\bibinfo
  {journal} {\apj}\ }\textbf {\bibinfo {volume} {919}},\ \bibinfo {eid} {53}
  (\bibinfo {year} {2021})},\ \Eprint {https://arxiv.org/abs/2105.00120}
  {arXiv:2105.00120} \BibitemShut {NoStop}%
\bibitem [{\citenamefont {{Brandenburg}}\ \emph {et~al.}(2019)\citenamefont
  {{Brandenburg}}, \citenamefont {{Bracco}}, \citenamefont {{Kahniashvili}},
  \citenamefont {{Mandal}}, \citenamefont {{Roper Pol}}, \citenamefont
  {{Petrie}},\ and\ \citenamefont {{Singh}}}]{brandenburg......19}%
  \BibitemOpen
  \bibfield  {author} {\bibinfo {author} {\bibfnamefont {A.}~\bibnamefont
  {{Brandenburg}}}, \bibinfo {author} {\bibfnamefont {A.}~\bibnamefont
  {{Bracco}}}, \bibinfo {author} {\bibfnamefont {T.}~\bibnamefont
  {{Kahniashvili}}}, \bibinfo {author} {\bibfnamefont {S.}~\bibnamefont
  {{Mandal}}}, \bibinfo {author} {\bibfnamefont {A.}~\bibnamefont {{Roper
  Pol}}}, \bibinfo {author} {\bibfnamefont {G.~J.~D.}\ \bibnamefont
  {{Petrie}}},\ and\ \bibinfo {author} {\bibfnamefont {N.~K.}\ \bibnamefont
  {{Singh}}},\ }\bibfield  {title} {\bibinfo {title} {{E and B Polarizations
  from Inhomogeneous and Solar Surface Turbulence}},\ }\href
  {https://doi.org/10.3847/1538-4357/aaf383} {\bibfield  {journal} {\bibinfo
  {journal} {\apj}\ }\textbf {\bibinfo {volume} {870}},\ \bibinfo {eid} {87}
  (\bibinfo {year} {2019})},\ \Eprint {https://arxiv.org/abs/1807.11457}
  {arXiv:1807.11457} \BibitemShut {NoStop}%
\bibitem [{\citenamefont {{Mignone}}(2014)}]{mignone14}%
  \BibitemOpen
  \bibfield  {author} {\bibinfo {author} {\bibfnamefont {A.}~\bibnamefont
  {{Mignone}}},\ }\bibfield  {title} {\bibinfo {title} {{High-order
  conservative reconstruction schemes for finite volume methods in cylindrical
  and spherical coordinates}},\ }\href
  {https://doi.org/10.1016/j.jcp.2014.04.001} {\bibfield  {journal} {\bibinfo
  {journal} {\jcph}\ }\textbf {\bibinfo {volume} {270}},\ \bibinfo {pages}
  {784} (\bibinfo {year} {2014})},\ \Eprint {https://arxiv.org/abs/1404.0537}
  {arXiv:1404.0537} \BibitemShut {NoStop}%
\bibitem [{\citenamefont {{Ustyugov}}\ \emph {et~al.}(2009)\citenamefont
  {{Ustyugov}}, \citenamefont {{Popov}}, \citenamefont {{Kritsuk}},\ and\
  \citenamefont {{Norman}}}]{ustyugov...09}%
  \BibitemOpen
  \bibfield  {author} {\bibinfo {author} {\bibfnamefont {S.~D.}\ \bibnamefont
  {{Ustyugov}}}, \bibinfo {author} {\bibfnamefont {M.~V.}\ \bibnamefont
  {{Popov}}}, \bibinfo {author} {\bibfnamefont {A.~G.}\ \bibnamefont
  {{Kritsuk}}},\ and\ \bibinfo {author} {\bibfnamefont {M.~L.}\ \bibnamefont
  {{Norman}}},\ }\bibfield  {title} {\bibinfo {title} {{Piecewise parabolic
  method on a local stencil for magnetized supersonic turbulence simulation}},\
  }\href {https://doi.org/10.1016/j.jcp.2009.07.007} {\bibfield  {journal}
  {\bibinfo  {journal} {\jcph}\ }\textbf {\bibinfo {volume} {228}},\ \bibinfo
  {pages} {7614} (\bibinfo {year} {2009})},\ \Eprint
  {https://arxiv.org/abs/0905.2960} {arXiv:0905.2960}
  \BibitemShut {NoStop}%
\bibitem [{\citenamefont {{Falkovich}}(1994)}]{falkovich94}%
  \BibitemOpen
  \bibfield  {author} {\bibinfo {author} {\bibfnamefont {G.}~\bibnamefont
  {{Falkovich}}},\ }\bibfield  {title} {\bibinfo {title} {{Bottleneck
  phenomenon in developed turbulence}},\ }\href
  {https://doi.org/10.1063/1.868255} {\bibfield  {journal} {\bibinfo  {journal}
  {Physics of Fluids}\ }\textbf {\bibinfo {volume} {6}},\ \bibinfo {pages}
  {1411} (\bibinfo {year} {1994})}\BibitemShut {NoStop}%
\bibitem [{\citenamefont {{Monin}}\ and\ \citenamefont
  {{Yaglom}}(1965)}]{monin.65}%
  \BibitemOpen
  \bibfield  {author} {\bibinfo {author} {\bibfnamefont {A.~S.}\ \bibnamefont
  {{Monin}}}\ and\ \bibinfo {author} {\bibfnamefont {A.~M.}\ \bibnamefont
  {{Yaglom}}},\ }\bibfield  {title} {\bibinfo {title} {{\em Statistical Fluid
  Mechanics: The Mechanics of Turbulence}}}, Nauka, Moscow \href@noop {} {\ \textbf
   (\bibinfo {year} {1965}), Vol. 1}, Ch.~2, \S3.3 and \S4.7\BibitemShut {NoStop}%
\end{thebibliography}


\end{document}